\documentclass[a4paper,11pt,notitlepage]{article}
\usepackage[margin=1in]{geometry}
\usepackage{amsmath}
\usepackage{amssymb}
\usepackage{natbib}
\usepackage{xcolor}
\usepackage{dsfont}
\usepackage{tikz}
\usepackage{graphicx}
\usepackage{subcaption}
\usepackage{authblk}
\newcommand*\diff{\mathop{}\!\mathrm{d}}
\providecommand{\keywords}[1]
{
  \small	
  \textbf{\textit{Keywords---}} #1
}
\title{A Bayesian \textit{block maxima over threshold} approach applied to corrosion assessment in heat exchanger tubes}

\author{Jess Spearing, Jarno Hartog}
\affil{\footnotesize{\textit{Shell Global Solutions B.V., 1031 HW Amsterdam, The Netherlands.}}}

\begin{document}

\maketitle

\begin{abstract}
Corrosion poses a hurdle for numerous industrial processes, and though corrosion can be measured directly, statistical approaches are often required to either correct for measurement error or extrapolate estimates of corrosion severity where measurements are unavailable. This article considers corrosion in heat exchangers tubes, where corrosion is typically reported in terms of maximum pit depth per inspected tube, and only a small proportion of tubes are inspected, suggesting extreme value theory (EVT) as suitable methodology.
However, in data analysis of heat exchanger data, shallow tube-maxima pits often cannot be considered as extreme; although previous EVT approaches assume all the data are extreme. We overcome this by introducing a threshold --- suggesting a \textit{block maxima over threshold} approach, which leads to more robust inference around model parameters and predicted maximum pit depth.
The model parameters of the resulting left-censored generalized extreme value distribution are estimated using Bayesian inference, meaning parameter uncertainty is immediate from the estimation procedure. This framework also naturally extends to cases where there is significant measurement error or censoring of inspection results, which is a common challenge.

\end{abstract}

\section{Introduction}
\label{sec:introduction}
\begin{figure}[h!]
    \centering
    \includegraphics[width=0.5\linewidth]{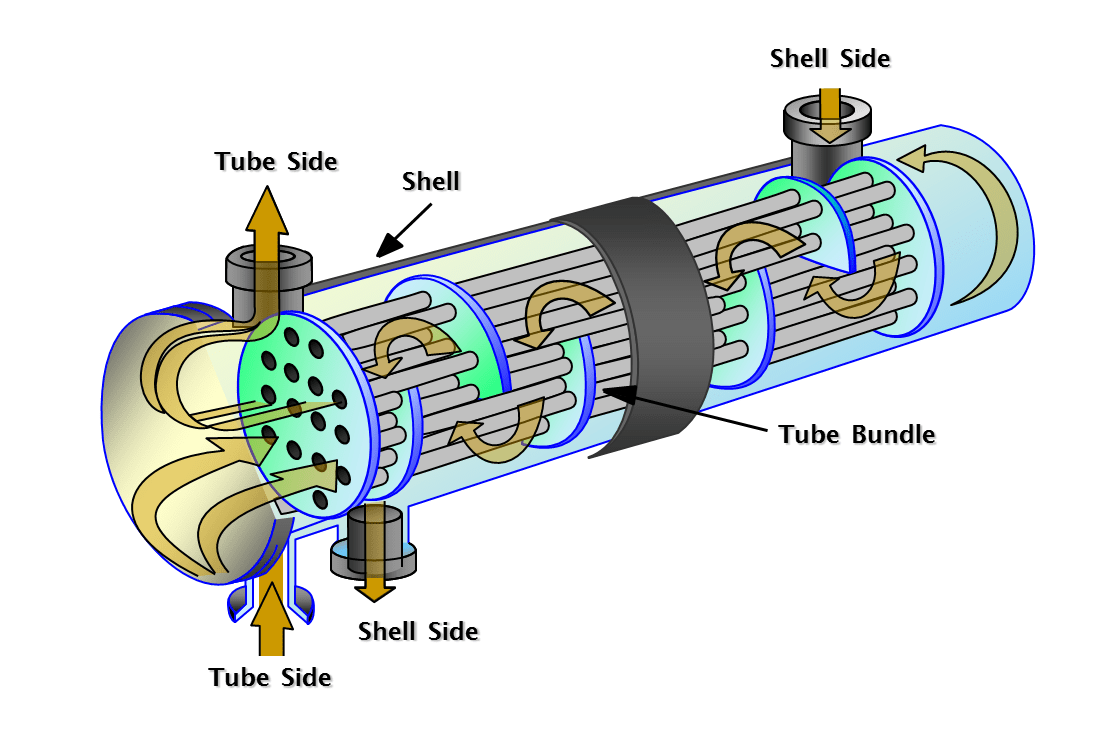}
    \caption{Diagram of a heat exchanger.}
    \label{fig:heat exchanger pic}
\end{figure}

Corrosion represents a significant challenge in various industries and affects the efficiency and integrity of critical equipment. Among the various corrosion mechanisms, pitting corrosion stands out as a particularly insidious threat due to its difficulty to detect and predict, potentially leading to unexpected and catastrophic failures. 
Figure \ref{fig:heat exchanger pic} depicts a heat exchanger, used in a wide range of applications, which heats or cools fluids.
The poisonous effects of pitting corrosion in heat exchangers are two-fold. Firstly, it compromises the structural integrity of the equipment. The localized nature of pitting corrosion creates small holes or pits in the metal surface, resulting in weakened areas susceptible to leakage and eventual rupture. Secondly, it impacts the thermal performance of a heat exchanger. The presence of pits disrupts the smooth flow of fluids within the heat exchanger, leading to reduced heat transfer efficiency and increased energy consumption.
This article focuses on the former, with the overall goal of assessing the likelihood of a leak/rupture in order to aid decision making around replacing of tubes.


Detection and monitoring of pitting corrosion in heat exchangers is a well-established practice, with traditional inspection methods such as visual examination, ultrasonic testing, and eddy current testing offering valuable insights into the condition of visible surfaces. Despite this, it is usually impractical or infeasible to reliably inspect all parts of the heat exchanger --- either due to cost or that some areas of the heat exchanger may be inaccessible --- and therefore the available measurements cannot accurately depict the state of the heat exchanger \citep{yu2016micromechanics}. With these difficulties and high costs of inspection, a statistical analysis is required, which predicts the likely corrosion depth on those uninspected tubes, given the available data from the observed tubes. 

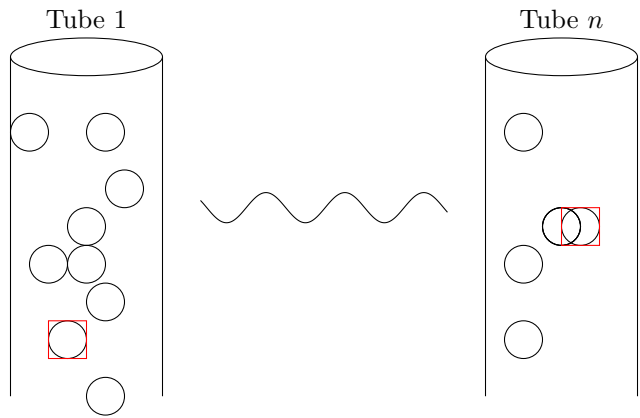
\begin{figure}[ht!]
    \centering
\begin{tikzpicture}
\draw  (4,21.75) ellipse (1cm and 0.25cm);
\draw (3,21.75) -- (3,17.25);
\draw (5,21.75) -- (5,17.25);
\draw  (3.25,20.75) circle (0.25cm);
\draw  (4,19.5) circle (0.25cm);
\draw  (4.25,18.5) circle (0.25cm);
\draw  (3.75,18) circle (0.25cm);
\draw  (3.5,19) circle (0.25cm);
\draw  (4,19) circle (0.25cm);
\draw  (4.25,20.75) circle (0.25cm);
\draw  (4.25,17.25) circle (0.25cm);
\draw [ color={rgb,255:red,255; green,0; blue,0} ] (3.5,18.25) rectangle (4,17.75);
\draw  (10.25,21.75) ellipse (1cm and 0.25cm);
\draw (9.25,21.75) -- (9.25,17.25);
\draw (11.25,21.75) -- (11.25,17.25);
\draw  (10.25,19.5) circle (0.25cm);
\draw  (10.25,19.5) circle (0.25cm);
\draw  (10.25,19.5) circle (0.25cm);
\draw  (9.75,18) circle (0.25cm);
\draw  (10.5,19.5) circle (0.25cm);
\draw  (9.75,20.75) circle (0.25cm);
\draw  (9.75,19) circle (0.25cm);
\draw [ color={rgb,255:red,255; green,0; blue,0} ] (10.25,19.75) rectangle (10.75,19.25);
\node [font=\small] at (3.25,20.75) {};
\node [font=\small] at (4.25,17.25) {};
\node [font=\small] at (9.75,18) {};
\node [font=\small] at (4,22.25) {Tube 1};
\node [font=\small] at (10.25,22.25) {Tube $n$};
\draw  (4.5,20) circle (0.25cm);
\node [font=\small] at (3.75,18) {};
\node [font=\small] at (11,19.5) {};
\draw[domain=5.5:8.75,samples=100,smooth] plot (\x,{0.2*sin(6.05*\x r -5.5 r ) +19.75});
\end{tikzpicture}
\caption{Sketch of a heat exchanger with pitting corrosion. The black circles indicate the pitting corrosion for a particular tube, and the red squares indicate the most extreme corrosion from each tube. Typically the observed data are only the pit depths of the most extreme corrosion per tube (the red squares), giving a one-dimensional dataset of size $n$.}
    \label{fig:paint sketch}
\end{figure}

Figure~\ref{fig:paint sketch} depicts a sketch of pit depth data on two (generalized to $n$) tubes within a heat exchanger. Each tube has multiple `pits' --- small holes --- in the surface, their locations indicated by black circles, with red squares indicating the deepest pit from each tube. 
There are three main practical issues with collecting such data: (i) it is often difficult to decipher a single pit from multiple pits when they are close together on a tube, e.g., the deepest pit (red square) on tube $n$ in Figure \ref{fig:paint sketch}; (ii) shallow pits are generally hard to measure and lead to large measurement uncertainty; (iii) within pitting corrosion there are different classes of pits, e.g., some pits initiate and then are quickly stifled, whilst others continue to deepen perpetually \citep{aziz1956application}. Those that deepen perpetually are the data of interest, but this distinction is difficult to observe. 
In the context of rupture risk due to pitting corrosion, typical maintenance involves plugging or replacing the the entire tube, and so the condition of a heat exchanger tube is characterized by its deepest pit. Therefore typical statistical analyses seek to identify the likely \textit{maximum} pit depth across all unobserved tubes. If this value is too large, then more tubes must be inspected, or maintenance is required.

Addressing questions around ``maxima'' or ``severe'' naturally lead the statistical methodology into the field of extreme value theory (EVT), which provides a framework for analysing rare events, and moreover, for specifically analysing maxima (here maximum pit depth). Indeed, The American Society for Testing and Materials' (ATSM) recommended standard practice for statistical analysis of pitting corrosion \citep{g46} has its roots in EVT. The recommendation is motivated by the work of~\cite{aziz1956application}, which analysed pitting corrosion on a collection of five aluminium alloys, using the maximum pit depth as the variable of interest.
It might seem wasteful to only consider the maximum pit depth per tube, when surely all deep pits could be useful for our analysis, especially as the distribution of the number of pits per tube has already been shown to be Poisson \citep{mears1937corrosion}. However, in addition to the data of all deep pits often not being available, analysing the tube maxima deals with issue (i), i.e., it is not always clear what is a single large pit versus many small pits when they are close together, and also helps with an unknown dependence structure between pit depths along the same tube. 


The specific recommended EVT approach of the ATSM mentioned above is to use a Gumbel distribution for tube-maxima pit depths. However, this Gumbel assumption used by~\cite{aziz1956application} contradicts the work of \cite{babu2019analysis}, which tested a range of distributions to model pitting corrosion across different materials and found that none of the experimental data could be shown to follow a Gumbel distribution; although, it could all be shown to follow a Weibull distribution. This finding motivates a more flexible distribution for maxima that can incorporate a variety of tail behaviours. Indeed more recent approaches \citep{lay2016} generalize the distribution of maxima to a generalized extreme value (GEV) distribution. The GEV assumption has a strong asymptotic justification as well as incorporates the findings of both \cite{aziz1956application} and \cite{babu2019analysis} since both Gumbel and Weibull distribution are special cases of the GEV, and the analysis of tube-maxima remedies issue (i). However, the applying the GEV approach to tube-maxima pit depths is not guaranteed to solve issue (ii). That is, the maximum pit depths per tube are not guaranteed to be large, and thus may contain large measurement uncertainty. Furthermore, there is no guarantee that the tube-maxima pit depths are large enough to be considered `extreme', which is one of the requirements for the GEV's asymptotic justifications. Moreover, differentiating between different classes of pits (issue iii) is more difficult when the pit depths are shallow, since a higher proportion of the deeper pits are likely to be those that deepen perpetually.

We attempt a remedy for issues (ii-iii) by only considering the deepest pits \textit{above a certain threshold}. The additional thresholding approach to the standard GEV serves to protect against modelling: data with large measurement error; model misspecification where the data cannot be considered `extreme'; and modelling multiple classes of pits that might otherwise give rise to a mixture distribution for pit depth (violating any single distributional assumption). 

The ensuing approach, termed the \textit{block maxima over threshold} (BMOT), generalizes the extreme value approach and results in a \textit{censored} generalized extreme value distribution with some constraints on its parameters. The paper presents the model within a Bayesian framework so gives us uncertainty quantification as well as parameter estimates. This is particularly useful since the frequentist approach of bootstrap techniques in extreme value applications can lead to nonsensical inference \citep{bickel1981some}. Moreover, it naturally extends to cases in which the data is censored or if there is significant measurement error through Bayesian imputation \citep{little2019statistical}. 

The paper is structured as follows. We first detail the background theory of univariate extreme value theory in Section~\ref{sec:theory}. In Section~\ref{sec:Model} we present the BMOT approach and relate it to current approaches such as \cite{g46} and \cite{lay2016}. Section~\ref{sec:Inference} presents the Bayesian framework for inference, after which we demonstrate its properties in simulated and real-world applications, Sections~\ref{sec:sim study} and \ref{sec:real data}, respectively.

\section{Theory}
 \label{sec:theory}
\subsection{Block maxima}
\label{sec:appendix gev}
As eluded to in Section \ref{sec:introduction}, extreme value theory (EVT) underpins our statistical analysis of extreme events. EVT arises fundamentally as the limiting behaviour of the maximum of a random sample, and in its simplest form pertains to an independent identically distributed (IID) random sample $X_1,\dots, X_k$ with each variable having a continuous distribution function $F$. Two common approaches in EVT are the block maxima (BM) method and the peaks over threshold (POT) methods, and the asymptotic theory is as follows. Let $M_k= \max\{X_1,\dots, X_k\}$ be the maximum of a block of length $k$, then we desire the distribution of $M_k$ for large $k$, and in particular appropriate choices of norming sequences $a_k > 0$ and $b_k$, such that, as $n\rightarrow\infty$,
\begin{eqnarray}
\Pr\left\{\frac{M_k-b_k}{a_k} \leq x \right\} & = & \Pr(X_1 \leq a_k x +
 b_k, \dots, X_k \leq a_k x+b_k) \nonumber \\
 & = & F^n(a_k x + b_k) \nonumber \\
 & \rightarrow & G(x)
\label{eq:probMax}
\end{eqnarray}
with non-degenerate limiting distrubtion $G(x)$. It turns out that the only possible non-degenerate limiting distribution of equation~\eqref{eq:probMax} is the generalised extreme value (GEV) distribution function, with the form
\begin{equation}
G(x) = \exp\left(-[1+\xi(x-\mu)/\sigma]^{-1/\xi}_+\right),
\label{eq:GEV}
\end{equation} 
 where $\mu \;\in \mathbb{R}, \;\sigma \in \mathbb{R}^+, \; \xi \in \mathbb{R}$, are the location, scale and shape parameters, respectively and $ z_+ = \max(z,0)$. For many applications, $X_1,\dots,X_k$ are likely to have some small dependence; however, the sizes of the blocks are taken to be large enough so that independence can be assumed between blocks. 

Whereas $\mu$ and $\sigma$ are simply determined by the scale of the data and therefore change based on, e.g., the measurement scale such as Kelvin vs. Celcius, the shape parameter $\xi$ is invariant to linear changes in measurement scale and is therefore dictated by the underlying physical process that generated the data, and determines the type of the distribution. A shape parameter $\xi<0$ corresponds to a Weibull distribution and as such there is a finite upper end-point to the distribution $x_G = \mu - \sigma/\xi: \; G(x) = 1, \; \forall x>x_G $. In contrast, for $\xi=0$ or $\xi>0$, corresponding to a Gumbel or Frechet distribution, respectively, then $G(x) < 1, \; \forall x<\infty$, i.e., there is no upper end-point to the distribution. The GEV result is powerful: it holds as the limit distribution for a very broad class of continuous distributions $F$ and implies that whatever $F$ is in this class, the maxima must follow a single class of distributions, determined by only three parameters.

The BM method \citep{coles2001introduction} assumes that, for a large enough block size $k$, limit~\eqref{eq:probMax} holds exactly. In many applications, given a sample of length $nk$ the approach is to split the series into $n$ blocks with $k$ values in each block. Often, however, this split arises naturally, for example in a heat exchanger with $n$ being the number of tubes and then $k$ being the number of deep pits on each tube (which can be different for each tube). Then the $n$ values of the block maxima are used to estimate the parameters $(\mu, \sigma, \xi)$ of the model, assuming that each of these variables is IID and follows a GEV.

Taking the maxima can help with cases when observations are not IID \textit{within} a block, but maxima \textit{across} blocks can be treated as IID. Pitting corrosion in heat exchanger tubes are a good example of this, and so motivates the use of the GEV. However, in heat exchanger data the maximum pit depth per tube is provided \textit{without} knowing the total number of pits per tube. That is, we do not know the block size, so cannot guarantee whether $k$ is `sufficiently large' enough to approximate the limit. It is for this reason we must also consider the peaks over threshold approach.


\subsection{Peaks over threshold}
\label{sec:appendix gpd}
The peaks over threshold (POT) approach considers all of the large values, rather than only the maxima. Though this may appear irrelevant in our application --- as heat exchanger data typically pertain to tube-maxima pit depths --- we will see later that the asymptotic theory it provides is still useful in the setting of maxima.
Let \[N_k(x) = \sum_{i=1}^k \mathds{1}\left( X_i > a_k x + b_k \right), \] with $\mathds{1}(A)$ be an indicator of event $A$ occurring, then $N_k(x)$ is the random variable corresponding to the number of $X_1,\dots,X_k$ exceeding the threshold $a_k x + b_k$, with $a_k$ and $b_k$ given in limit~\eqref{eq:probMax}. Therefore $N_k(x)$ has a Binomial distribution with $N_k(x) \sim \operatorname{B}(n, 1 - F(a_k x + b_k))$. 
From the same GEV limit conditions of equation~\eqref{eq:probMax}, as $k \rightarrow\infty$,
\begin{equation*}
k \log F(a_k x + b_k) \rightarrow \log G(x).
\end{equation*}
Therefore, using standard Taylor series approximation, for all $x$
 \begin{equation}
 \label{eq:gpd2}
 k[1-F(a_k x + b_k)] \rightarrow -\log G(x) = [1+\xi(x-\mu)/\sigma]^{-1/\xi}_+.
 \end{equation}
From property~\eqref{eq:gpd2}, then the standard Poisson limit from a Binomial gives that as $k~\rightarrow~\infty$, $N_k( x )~\rightarrow~N(x)$, where $N(x)$ is a Poisson random variable with mean $[1+~\xi(x-~\mu)/\sigma]^{-1/~\xi}_+$. In addition, it follows that for $x>u$ and $X$ distributed as $X_i$, we have that
\begin{equation}
\Pr\{X > a_k x + b_k | X > a_k u + b_k\} \rightarrow \log G(x)/ \log G(u) =: \bar{H}_u(x),
\label{eq:probExc}
\end{equation}
where $\bar{H}_u(x) := 1-H_u(x)$, and where the distribution function $H_u$ is of the form
\begin{equation}
H_u(x) := 1-\left[1+\xi \left(\frac{x-u}{\tilde{\sigma}_u}\right)\right]^{-\frac{1}{\xi}}_+,
\label{eq:GPd}
\end{equation}
is the generalised Pareto distribution function (GPD) with: threshold $u$; shape parameter $\xi$, which is the same as that from the GEV; and scale parameter $\tilde{\sigma}_u \in \mathbb{R}^+$, which is linked to the GEV parameters via $\tilde{\sigma}_u = \sigma + \xi(u-\mu)$. The limit distribution $H_u$ provides an asymptotic model for the distribution of exceedances above a threshold $u$, no matter what the distribution $F$. 
Similarly to the GEV, for $\xi < 0$, there exists a finite value $x_H = u - \tilde{\sigma}_u/\xi: \; H_u(x) = 1, \; \forall x>x_H $. In contrast, for $\xi\geq 0, \; H_u(x) < 1, \; \forall x<\infty$. The POT approach therefore leads to a model for the extreme tail with two components: a model for the number of exceedances of the threshold, which is Poisson with mean $\lambda = [1+\xi(x-\mu)/\sigma]^{-1/\xi}_+$, and a model for threshold exceedances, $H_u(x)$ which is GPD. The choice of threshold $u$ is user-specified, with the choice based on the usual bias-variance trade-off, the subject of much historical focus \citep{scarrott2012review}.

In heat exchanger data where the tube-maxima is only known, applying standard POT methodology is not suitable since all the data over a high threshold are not available. However, by combining the POT methodology with a BM approach we will demonstrate how to apply the block maxima approach when the block size is unknown (and potentially small), by thresholding the maxima, in a methodology we term \textit{Block maxima over threshold} (BMOT).


\section{Modelling: block maxima over threshold}
\label{sec:Model}
\subsection{Generative distribution for pit depths}
Although we only use data of tube maxima pit depths for inference, the justification of the distribution of tube-maxima is derived from the assumption for how \textit{all} deep pits form. To begin, we assume that the number of pits per tube is IID across different tubes in the bundle. Moreover, the distribution of pit depth --- conditional on the number of pits --- is also IID across tubes. In practice, some tubes might be more prone to corrosion, e.g., due to the geometry or environment of the heat exchanger. This is discussed in Section~\ref{sec:Discussion}.

The number of these pits per tube can be shown to be Poisson distribution as in \cite{mears1937corrosion}, i.e., the number of pits $K$ has distribution $K\sim \text{Poisson}(\lambda_0)$, for some rate $\lambda_0>0$. Note that if each tube
has the same surface area, then $\lambda_0$ is constant over different tubes. We assume this throughout
the rest of this paper, although the results generalize to non-constant $\lambda_0$. 



The assumed underlying distribution of all pit depth occurrences on a tube here is modelled via a generalized Pareto distribution (GPD) with shape and scale parameters $\xi \in \mathbb{R}$ and $\tilde{\sigma}_0\in\mathbb{R}_+$, respectively. The GPD is a broad class of tail distributions, for example, it encompasses the exponential distribution as a special case, and so allows for the fact that the distribution of pit depths depends on the metal and the surrounding environmental \citep{akpanyung2019pitting}. Moreover, the GPD is an extreme value distribution and as such has strong asymptotic justifications as observed in Section \ref{sec:theory}.
Then, the probability the tube contains $k\geq 0$ pits is $\Pr(K=k) = \frac{\lambda_0^k e^{-\lambda_0}}{k!}$, and if a pit depth $X$ is modelled as $X\sim \text{GPD}(\tilde{\sigma}_0, \xi)$, then $\Pr(X>x) = \left(1 + \xi \frac{x}{\tilde{\sigma}_0}\right)_+^{-1/\xi}$ for $x > 0$. 

Conditional on observing the number of pits from a single tube, the pit depths themselves are assumed independent. The distribution of maximum pit depth, $Y:=\max\{X_1, \dots, X_{K}\}$, where $X_k$ represents the $k$th pit depth, on a single tube becomes
\begin{align}
    \Pr(Y\leq x) &= \sum_{k=0}^{\infty} \Pr(\max\{X_1,\dots, X_{K}\} < x | K = k)\Pr(K=k)\nonumber\\
     &= \sum_{k=0}^{\infty}  \left(1-\left(1 + \xi x/\tilde{\sigma}_0\right)_+^{-1/\xi}\right)^k \frac{\lambda_0^k e^{-\lambda_0}}{k!} \nonumber\\
     &= e^{-\lambda_0}\sum_{k=0}^{\infty}\frac{\left(\lambda_0 \tau(x)\right)^k}{k!} = e^{-\lambda_0}e^{\lambda_0 \tau(x)}\nonumber\\
     &=\exp\left\{-\lambda_0(1 + \xi x/\tilde{\sigma}_0)_+^{-1/\xi}\right\},
     \label{eq:maximum pit depth derivation}
\end{align}
where $K$ is the (random) count of pit depth observations on a given tube, and $\tau(x):=1-\left(1 + \xi x/\tilde{\sigma}_0\right)_+^{-1/\xi}$. The probability \eqref{eq:maximum pit depth derivation} has the form of the Generalised extreme value distribution, of which both Gumbel and Weibull are a special case \citep{coles2001introduction}.

\subsection{Thresholding}
In practice however, small pits are difficult to measure, or could also arise from some unrelated form of corrosion, i.e., not pitting corrosion, that we do not wish to model, and current approaches don't attempt to account for this. Furthermore, for the asymptotic justification to hold, the GPD requires that pit depths are sufficiently deep. Therefore it is prudent to only consider deeper pits that are greater than some level, or \textit{threshold}, say $u>0$. 
This provides the flexibility to deal with cases where the GPD assumption for pit depth is misspecified, because
we know from limit theory \citep{coles2001introduction} that the GPD is the only limit distribution for exceedances as the threshold increases. Therefore, if it is found that the model gives a poor fit for a selected $u$, the threshold can be raised to some $v>u$ until we observe stability, subject to the usual bias-variance trade-off. 
In short, even if the generative model is misspecified, the introduction of a threshold --- and its accompanying limit theory --- mitigates the risk of significantly under or overestimating corrosion pit depth. This is the main benefit of this approach over a standard GEV fit without use of a threshold.

We therefore model $X-u|X>u \sim \text{GPD}(\tilde{\sigma}_u, \xi)$, i.e., pit depths greater than $u>0$ follow a GPD, where $K_u\sim \text{Poisson}(\lambda_u)$ is now the number of pits with pit depth greater than $u$, and $0<\lambda_u\leq \lambda_0$, and $\tilde{\sigma}_u = \tilde{\sigma}_0 + \xi u$. The remaining observations $X|X\leq u$ are considered censored to $u$. 
It follows that the deepest pit per tube $Y$ has a \emph{censored} GEV distribution


\begin{equation}
    \label{eq:censored distribution}
        \Pr(Y\leq x) = 
                \exp\left\{-\lambda_u\left[1 + \xi\left(\frac{x-u}{\tilde{\sigma}_u}\right)\right]_+^{-1/\xi}\right\}
        \end{equation}
for $x\geq u$, and which is GEV with shape $\xi$, location $\mu = u + \tilde{\sigma}_u (\lambda_u^{\xi}-1) / \xi$ and scale $\sigma = \tilde{\sigma}_u \lambda_u^\xi$, left-censored at threshold $u$.



The censored-GEV is derived as follows. The probability of no deep pits in our model is $P(K=0) = e^{-\lambda_u}$ and corresponds to the case $x=u$ in Equation \eqref{eq:censored distribution}. To verify the censored-GEV parameters, note that

\begin{equation*}
\begin{split}
\left(1 + \xi \frac{x - \mu}{\sigma}\right)_+^{-1/\xi} & = \left(1 + \xi \frac{x - u - \tilde{\sigma}_u(\lambda_u^\xi - 1)/\xi}{\tilde{\sigma}_u\lambda_u^\xi}\right)_+^{-1/\xi} \\
 & = \left(1 + \lambda_u^{-\xi}\xi \frac{x-u}{\tilde{\sigma}_u} - 1 + \lambda_u^{-\xi}\right)_+^{-1/\xi} \\
 & = \lambda_u\left(1+\xi \frac{x-u}{\tilde{\sigma}_u}\right)_+^{-1/\xi}.
\end{split}
\end{equation*}
Note that the original parameters $\xi, \tilde{\sigma}_u, \lambda_u$ can be recovered from the censored-GEV parameters $\xi, \mu, \sigma$ as 
\begin{equation}
\label{eq:paramters}
\begin{split}
\tilde{\sigma}_u &= \sigma - \xi (\mu - u), \\
\lambda_u &= \left(\frac{\sigma}{\sigma - \xi (\mu - u)}\right)^{1/\xi},
\end{split}
\end{equation}
with $\xi$ being the same in both parametrisations.
From this, we see that our generative model constrains the GEV parameter space by
\begin{equation*}
\sigma > \xi(\mu - u).
\end{equation*}
Note that the deepest pit depth in the bundle is again censored-GEV distributed by max-stability \citep{coles2001introduction}, since it is a maximum of left-censored GEV distributions. For a bundle of $n$ tubes, the distribution of the maximum pit depth has shape parameter $\xi$, location $u + \tilde{\sigma}_u (\lambda_u^{\xi}-1) / \xi + \tilde{\sigma}_u \lambda_u^\xi (n^\xi - 1) / \xi$ and scale $\tilde{\sigma}_u \lambda_u^\xi n^\xi$.
We can summarize our generative model with threshold $u$ as:
\begin{equation*}
\begin{split}
K|\lambda_u &\sim \text{Poisson}(\lambda_u) \\
X_1-u, \ldots, X_K-u|K, \xi, \tilde{\sigma}_u &\sim \text{GPD}(\tilde{\sigma}_u, \xi)
\end{split}
\end{equation*}
If we define $Y := \max\{X_1, \ldots, X_K\}$ for $K > 0$ and $Y = u$ for $K = 0$, then it follows that:
\begin{equation*}
\label{eq:distribution tube max}
Y | \xi, \tilde{\sigma}_u, \lambda_u \sim \text{GEV}(u + \tilde{\sigma}_u (\lambda_u^\xi - 1)/\xi, \tilde{\sigma}_u \lambda_u^\xi, \xi) \text{\ left-censored at $u$},
\end{equation*}
which we might also write as
\begin{equation}
Y | \mu, \sigma, \xi \sim \text{GEV}(\mu, \sigma, \xi) \text{\ left-censored at $u$}.
\label{eq:generative model}
\end{equation}

\subsection{Inference}\label{sec:Inference}
The generative model in equation~\eqref{eq:generative model} gives a distribution for the maximum pit depth on a single tube. The subset of tubes which are inspected provide the data which we use to infer the distribution of the maximum pit depths, assuming the tube maxima are iid. This can then be used to extrapolate
to the unobserved tubes. Typically, either the max-stability property is used to find the distribution of the maximum pit depth in the unobserved tubes, or maximum pit depth in
the unobserved tubes is estimated as a return level of the individual tube maximum pit depth distribution, such as in \cite{aziz1956application}. See \cite{phil2021} for a discussion and comparison of these approaches.

Let there a set of $\mathcal{N}$ observed tubes, of which $\mathcal{N}_+$ with $|\mathcal{N}_+|=:n_+$, and $\mathcal{N}_-$  with $|\mathcal{N}|_-=n$ are sets of tubes with maxima above and below the threshold, respectively. The contribution to the likelihood for set of tube maxima above the threshold $u$ are IID distributed with GEV$(\mu, \sigma, \xi)$, and for the observed maxima below the threshold, the contribution to the likelihood is $P(y\leq u) = e^{-\lambda_u}$.
Therefore, for tube-maxima pit depth data $\pmb{y}$ the likelihood is given as
\begin{equation*}
    L(\pmb{y}|\lambda_u, \sigma, \xi) =\exp(-\lambda_u)^{n-}\prod_{i\in\mathcal{N}_+} \left(\frac{1}{\sigma}\left[1 + \xi\left(\frac{Y_i-\mu}{\sigma}\right)\right]_+^{-1/\xi-1}\exp\left\{-\left[1 + \xi\left(\frac{Y_i-\mu}{\sigma}\right)\right]_+^{-1/\xi}\right\}\right)
\end{equation*}
The prior for our parameters $\theta:=(\lambda_u, \sigma, \xi)$ is assumed mutually independent across its components, i.e., $\pi_\theta(\theta) = \pi_\lambda(\lambda_u)\pi_\sigma(\sigma)\pi_\xi(\xi)$, with the marginal choices for these priors given as follows. The shape parameter prior $\pi_\xi(\xi)$ is given such that $\text{logit}(\xi+0.5)\sim \mathcal{N}(\text{logit}(0.5),\psi)$, for $\psi \in\mathbb{R}_+$ which restricts the domain of $\xi$ to be $\xi \in (-0.5,0.5)$ but is still differentiable across its whole domain for efficient gradient-based samplers. A Gamma prior is used for the scale parameter and rate of exceedances, such that $\sigma \sim \Gamma(\alpha_\sigma,\beta_\sigma)$ and $\lambda_u \sim \Gamma(\alpha_\lambda,\beta_\lambda)$ with $\alpha_\sigma,\beta_\sigma, \alpha_\lambda, \beta_\lambda\in\mathbb{R}_+$.
The hyper-parameters $\psi, \alpha_\sigma,\beta_\sigma, \alpha_\lambda, \beta_\lambda$ are selected based on the scale of the data and/or expert knowledge, see Sections~\ref{sec:sim study} and \ref{sec:real data}.
Then, the full posterior distribution is $\pi(\theta|\pmb{y}) \propto \pi_\theta(\theta) L(\pmb{y}|\lambda_u, \sigma, \xi)$. In our experiments below, Markov chain Monte Carlo (MCMC) was used for inference, specifically a No-U-Turn sampler (NUTS) \citep{hoffman2014no} via the Python package \textit{PyMC} \citep{abril2023pymc}. The Gelman-Rubin statistic \citep{gelman1992inference} was used as the convergence diagnostic.

\section{Simulation study}
\label{sec:sim study}
\begin{figure}[ht]
    \centering
    \begin{subfigure}{0.3\textwidth}
        \includegraphics[width=\textwidth]{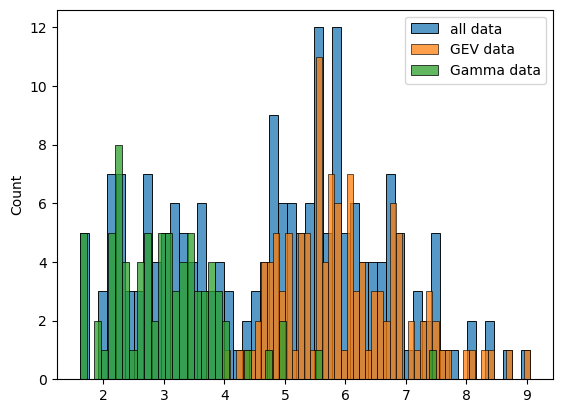}
        \caption{$\mu_\gamma=3$}
    \end{subfigure}
    \hfill
    \begin{subfigure}{0.3\textwidth}
        \includegraphics[width=\textwidth]{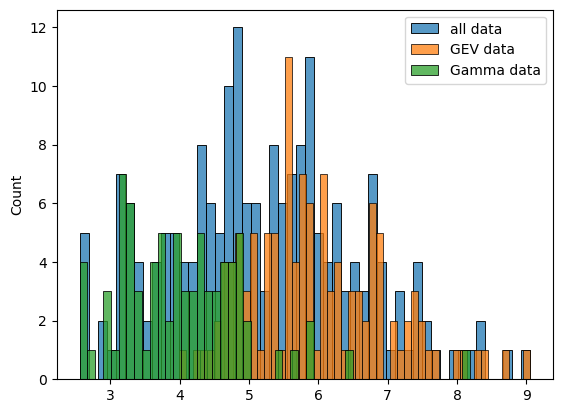}
        \caption{$\mu_\gamma=4$}
    \end{subfigure}
    \hfill
    \begin{subfigure}{0.3\textwidth}
        \includegraphics[width=\textwidth]{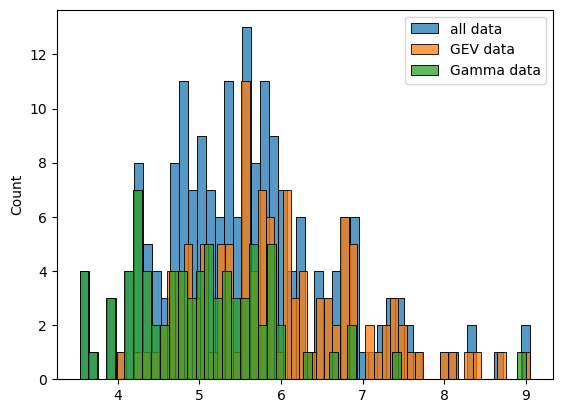}
        \caption{$\mu_\gamma=5$}
    \end{subfigure}
    \caption{Example of simulated data from the Gamma-GEV mixture distribution with $\xi=-0.1$, across the three different scenarios for the gamma mean.}
    \label{fig:data sim example}
\end{figure}

The value of introducing a threshold into extremes modelling of pitting corrosion data is to try to separate out the different classes of pits. Class (i) have an unknown distribution for pit depth, either because they: initiate and then are quickly stifled; or have simply not deepened enough to be determined `deep' pits; or are too shallow to have reliable measurements taken. Class (ii) pits are those which are thought to have a GPD distribution.
We therefore test our model on simulated data from a mixture distribution, with one portion of the data being simulated from a GEV, and the other --- though in practice unknown --- is being simulated here from a Gamma distribution. The Gamma distribution was selected as the mixture component for the non-extremes because: it does not have a heavy tail like we might expect for deep pits; we are not required to select an upper end point and so can still simulate arbitrarily large values; it is bounded from below at 0 which is physically required to simulate pit depths.
To test the sensitivity to different possible conditions, we test 5 different values of the shape parameter $\xi$ for the GEV generated data $X_G$, such that $X_G \sim \text{GEV}(\mu, \sigma, \xi)$ with $\mu=5.5, \sigma=1$ and $\xi \in (-0.2, -0.1, 0.001, 0.1, 0.2)$, where $\xi=0.001$ gives effectively a Gumbel distribution.
To allow for different scenarios in the Gamma distributed data $X_\gamma$, we allow 3 different scenarios, such that $X_\gamma \sim \text{Gamma}(\alpha, \beta)$, with a fixed variance $\alpha/\beta^2=1$ and mean $\alpha/\beta :=\mu_\gamma\in (3, 4, 5)$. Scenario (a) illustrated in Figure \ref{fig:data sim example} represents a case where the distribution of class (i) and class (ii) pit depths are easily distinguishable, whereas in scenario (c) there is considerable overlap and distinction is more difficult. 

Figure~\ref{fig:data sim example} gives an example of the three different $X_\gamma$ scenarios for the case $\xi=-0.1$.
The total number of data points for each replication is $n=200$, which is a typical number of tubes to be inspected for a heat exchanger with about 1000-2000 total tubes \citep{lay2016}, and the ratio of the mixture components is also assumed random, such that the number of GEV data points, $N_G$, follows a binomial distribution with $N_G \sim \text{Bin}(n, p_G),$ with $p_G=0.6$. This was then repeated 100 times over different seeds, and with 3 scenarios for $\mu_\gamma$, and 5 parameter values for $\xi$, resulting in 1500 datasets.
For each of the 1500 simulated datasets of $n=200$ tubes, two candidate models were fitted: our censored-GEV model presented in Section~\ref{sec:Model} with a threshold of $u=5.5$; and a standard GEV model with no censoring. These were compared to a ``simulated fit'', that is, we fit a standard GEV model to only the GEV data (orange data in Figure~\ref{fig:data sim example}).
The two candidate models are measured by how closely their estimated GEV parameters and the ``factor 10'' return level compare to those from the simulated fit. Here, the factor 10 return level is simply the estimate of the worst case of corrosion over an unobserved $10n=2000$ tubes.
As the Bayesian paradigm gives distributions for the parameters for each of the 1500 simulations, they are summarised via the difference in means, i.e., bias, and the ratio of their standard deviation compared to the simulated fit.

\begin{figure}[ht]
    \centering
    \begin{subfigure}{0.45\textwidth}
        \includegraphics[width=\textwidth]{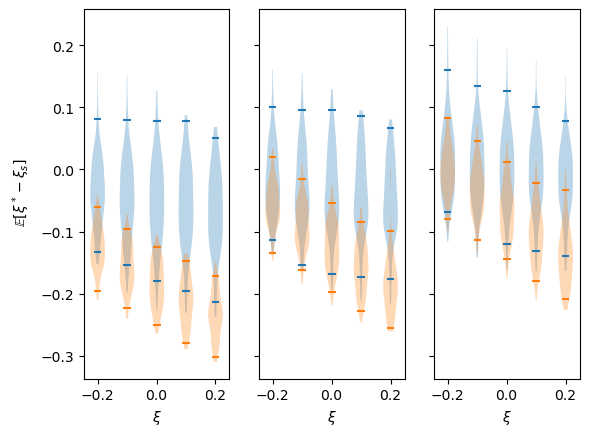}
        \caption{Bias}
    \end{subfigure}
    \hfill
    \begin{subfigure}{0.45\textwidth}
        \includegraphics[width=\textwidth]{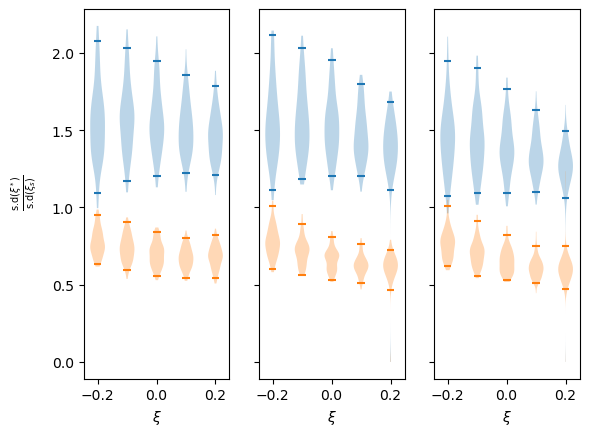}
        \caption{Std. dev. ratio}
    \end{subfigure}
    \caption{Posterior distribution of the shape parameter bias (a) and ratio of std. dev. (b) compared to the simulated model, with $95\%$ HDI bars, for our thresholded model (blue) and unthresholded model (orange).}
    \label{fig:sim results shape parameter}
\end{figure}

Here we investigate simulation results pertaining to the shape parameter and return levels, though those regarding the location and scale parameters can be seen in Appendix~\ref{sec: appendix additional sim figures}. Figure~\ref{fig:sim results shape parameter} shows the simulation results for the bias (a) and std. dev. ratio (b) of the shape parameter.
Compared to the true GEV data, our censored-GEV model (blue) shows no statistically significant bias in the shape parameter across all scenarios for $\xi$ and $\mu_\gamma$. The variance is on average approximately 1.5 times larger than simulated fit, inflating the uncertainty around this estimate.
For the standard GEV fit (orange) there are multiple instances of statistically significant negative bias in the shape parameter, meaning the model predicts a shorter tail than there is in reality and underestimates levels of corrosion. Not only this, but the ratio of the variance of the thresholded model is consistently smaller than 1, i.e., it is more confident than the true model, presumably because more data is used in the estimate. 

\begin{figure}[ht]
    \centering
    \begin{subfigure}{0.45\textwidth}
        \includegraphics[width=\textwidth]{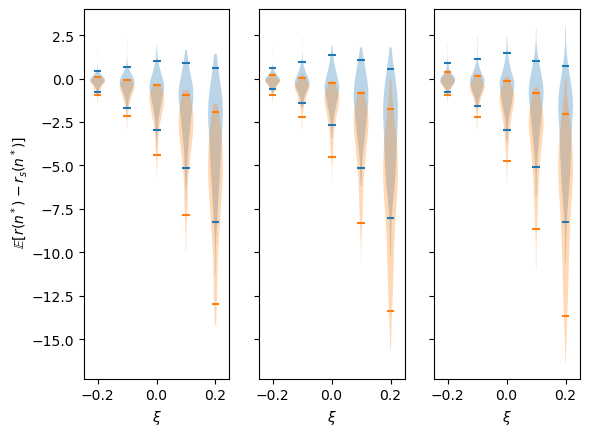}
        \caption{Bias}
    \end{subfigure}
    \hfill
    \begin{subfigure}{0.45\textwidth}
        \includegraphics[width=\textwidth]{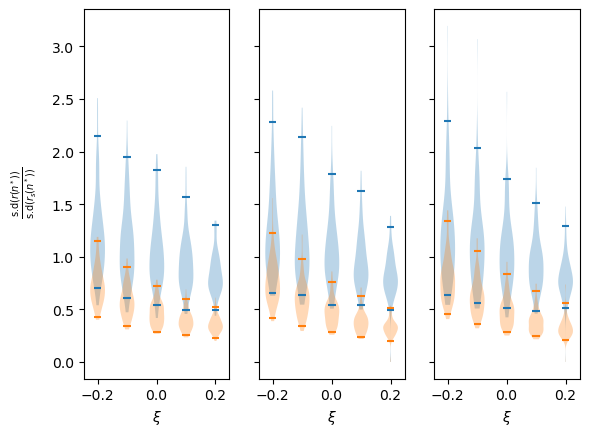}
        \caption{Std. dev. ratio}
    \end{subfigure}
    \caption{10-year return level: bias (a) and ratio of std. dev. (b) compared to the simulated model, for our thresholded model (blue) and unthresholded model (orange).}
    \label{fig:sim results return level}
\end{figure}

We see this more concretely in the return level estimates in Figure~\ref{fig:sim results return level} where the standard GEV model is generally underestimating levels of corrosion (a) and in many cases overconfidently so, i.e., the variance is smaller than baseline (b), with the effect becoming stronger as the true shape parameter $\xi$ of the simulated GEV data increases.
Our censored GEV model has no statistically significant bias in return levels, and the confidence is typically similar to the simulated model. That is, we typically recover the underlying simulated model's mean and variance of the return levels.

In this simulation therefore, we demonstrate our model's ability to recover good estimates for the return levels of corrosion data using extreme value theory, and which are conservative in their confidence intervals, even when the underlying data do not all follow an extreme value distribution.

\section{Real data example}
\label{sec:real data}
\subsection{Data}
The data are inspection data from a de-ethaniser condenser, consisting of 4834 tubes made out of A213 Grade 304L stainless steel.
Figure~\ref{sec:real data} displays the data as a histogram (a) and the empirical cdf (b). The heat exchanger in question comprises $n=367$ observations of tube-maxima corrosion, and $n^*=3489$ unobserved tubes, over which we wish to determine the maximum corrosion. The data are anonymized by a shift and linear scaling, and because of this the units of corrosion, or \textit{wall loss}, are not given here. The empirical cdf also shows that the data are rounded to +/- 0.01 on this scale. From these plots alone it is hard to assess the impact of any mixture distribution present. 
However, because we are interested in modelling the largest values, and therefore the upper tail, a censored-GEV helps to prevent potential model misspecification in the lower tail.
\begin{figure}[ht]
    \centering
    \begin{subfigure}{0.45\textwidth}
        \includegraphics[width=\textwidth]{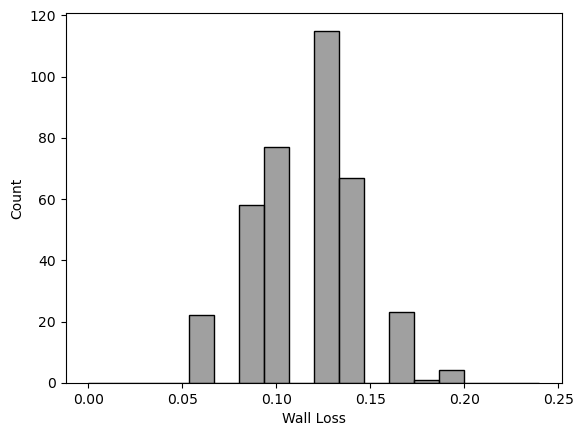}
        \caption{Histogram}
    \end{subfigure}
    \hfill
    \begin{subfigure}{0.45\textwidth}
        \includegraphics[width=\textwidth]{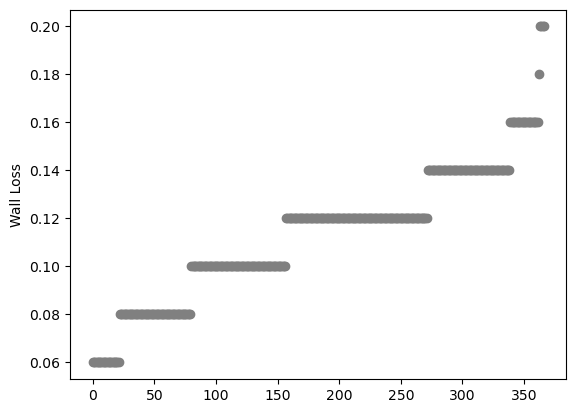}
        \caption{Empirical cumulative distribution function}
    \end{subfigure}
    \caption{Shifted and scaled wall loss data containing $n=367$ observations of tube-maximum corrosion.}
    \label{fig:real data}
\end{figure}

This rounding is compensated for in the analysis by treating it as interval-censored data and turning to Bayesian imputation. Each data point $y$ is considered a random variable $Y\in [y-0.01,y+0.01]$, and at each state in the MCMC procedure, a new $Y$ is drawn from an interval-censored GEV distribution.

\subsection{Prior selection}
The priors for our censored-GEV with parametrisation $(\lambda_u, \sigma, \xi)$ are of the form given in Section~\ref{sec:Inference}, with hyper parameters as follows. The rate parameter has prior $\lambda_u\sim\text{Gamma}(\alpha_\lambda, \beta_\lambda),$ with a mean which depends on the choice of threshold $\alpha_\lambda/\beta_\lambda=-\text{ln}(n_-/n)$ (where $n_-$ is the number of data below the selected threshold) and variance $\alpha_\lambda/\beta_\lambda^2=1$. With the selected threshold $u=0.11$ (see Section~\ref{sec:real data, threshold}) this gives $n_-/n=0.43$ and translates to a loosely informative prior for the rate of exceedances of the threshold with mean and 95\% highest density intervals (HDI) of $ 0.85 (0.006, 3.6)$. The scale parameter has prior $\sigma\sim\text{Gamma}(\alpha_\sigma, \beta_\sigma),$ with mean $\alpha_\sigma/\beta_\sigma=0.1$ and variance $\alpha_\sigma/\beta_\sigma^2=0.25,$ which again gives a loosely informative prior mean and 95\% HDIs of $ 0.1 (3.8\times 10^{-11}, 0.83)$. The standard deviation of the data is 0.03. The shape parameter has prior $\text{logit}(\xi+0.5)\sim \mathcal{N}(0,\psi)$, with $\psi=1.4$ which gives prior mean and 95\% HDI of $0 (-0.44, 0.44)$.

For the standard GEV model with parameters $(\mu, \sigma, \xi)$, the scale and shape parameters, $(\sigma,\xi)$ have the same priors as for the censored-GEV model, and the location parameter has the prior $\mu\sim\text{Normal}(0.1,1)$, which gives a 95\% HDI of $(-3.8, 4.0)$.

\subsection{Threshold selection}
\label{sec:real data, threshold}
\begin{figure}[ht!]
    \centering
    \includegraphics[width=0.5\linewidth]{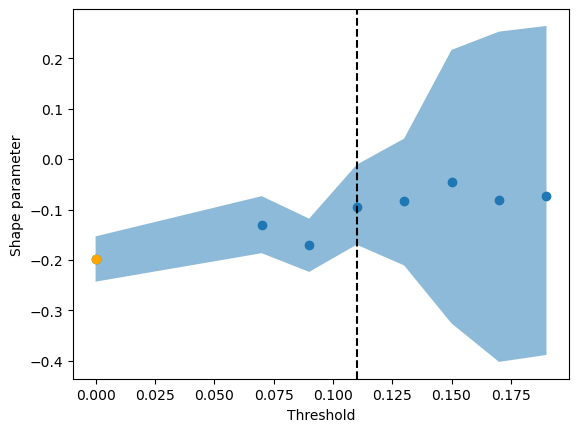}
    \caption{Threshold stability plot with points being the posterior mean of the shape parameter when fitted using the GEV model (orange) and the censored-GEV model (blue) for varying thresholds according to the x-axis. The shaded blue area represents 95\% HDIs. The vertical dashed line indicates the chosen threshold.}
    \label{fig:real data threshold selection}
\end{figure}

There are many methods of threshold selection, and this is the subject of much historical focus in univariate EVT, see \cite{scarrott2012review}. 
We opt for the threshold stability plot, see Figure \ref{fig:real data threshold selection}, which uses the fact that the shape parameter $\xi$ is invariant to the choice of threshold $u$. The censored-GEV model presented in Section \ref{sec:Model} is fit across a range of thresholds, specifically, at the upper/lower bounds of the rounding intervals, as well as a standard GEV model fit to all data (orange) placed at $u=0$. The plot shows posterior mean estimates of the shape parameter and 95\% HDIs (shaded region) for the different proposed thresholds $u$. We see stability in the estimate of the shape parameter from a threshold of $0.11$. For thresholds higher than this we see increases in the variance of the shape parameter estimate whilst no significant differences in its best estimate. Thus, subject to the usual bias-variance trade-off, the threshold is selected as $u=0.11$, which leaves $n_+=210$ of observations above the threshold.

\subsection{Model fit assessment}
\begin{figure}[ht]
    \centering
    \begin{subfigure}{0.3\textwidth}
        \includegraphics[width=\textwidth]{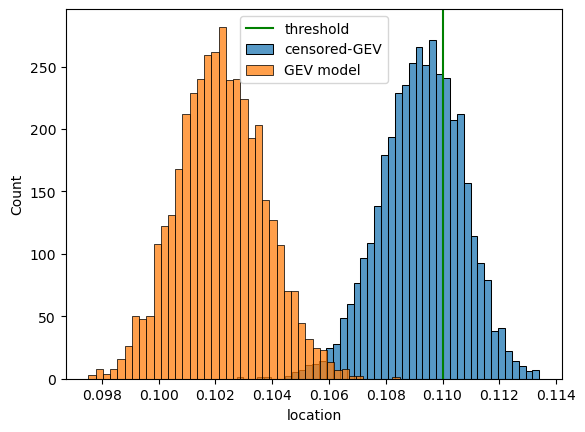}
        \caption{Location}
    \end{subfigure}
    \hfill
    \begin{subfigure}{0.3\textwidth}
        \includegraphics[width=\textwidth]{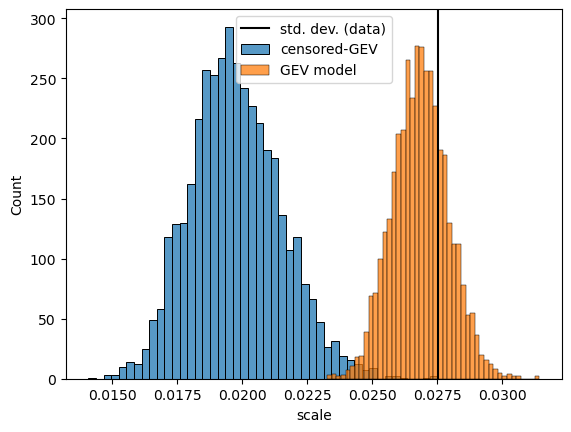}
        \caption{Scale}
    \end{subfigure}
    \hfill
    \begin{subfigure}{0.3\textwidth}
        \includegraphics[width=\textwidth]{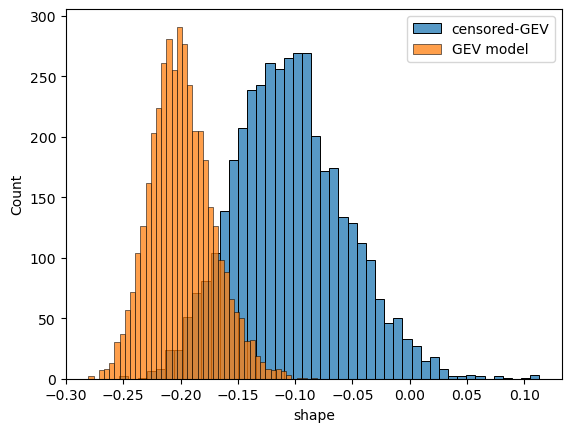}
        \caption{Shape}
    \end{subfigure}
    \caption{Location, scale and shape parameter fits of the standard GEV model (orange) and the censored-GEV model (blue), with $u=0.11$.}
    \label{fig:real data GEV parameters}
\end{figure}

Figure~\ref{fig:real data GEV parameters} shows the fitted location (a), scale (b), and shape (c) parameters for both the GEV model and our censored-GEV model, with threshold $u=0.11$.

The location parameter is significantly larger for the censored-GEV model (blue) than for the GEV model (orange), as now the median observation is larger due to the censoring of data below the threshold. The scale parameter for the censored-GEV model is significantly smaller than for the GEV model, presumably because the standard deviation of the data above the threshold is less than the full dataset. The shape parameter estimates are notably different, with the GEV model implying a shorter tail for the distribution of wall loss than the censored-GEV model, that is, the censored-GEV model gives an increased probability of more extreme wall loss.

\begin{figure}[h!]
    \centering
    \begin{subfigure}{0.48\textwidth}
        \includegraphics[width=\textwidth]{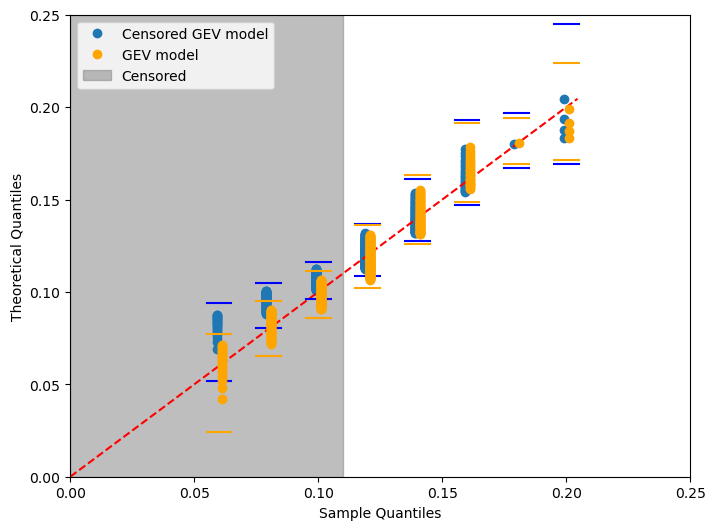}
        \caption{QQ-plot}
    \end{subfigure}
    \hfill
    \begin{subfigure}{0.48\textwidth}
        \includegraphics[width=\textwidth]{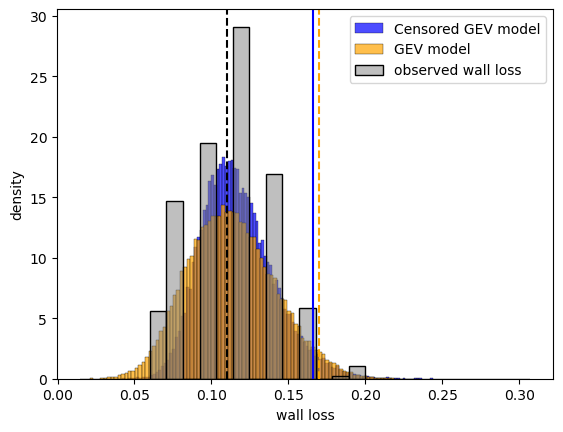}
        \caption{Posterior predictive}
    \end{subfigure}
    \caption{Two distinct methods of model fit assessment. The whiskers in the QQ-plot (a) indicate the respective models' 95\% HDI for each of rounding bins. The posterior predictive (b) indicates the models' upper 95\% HDI for the whole dataset. The black dashed line indicates the threshold $u$, the vertical yellow and blue lines are the upper 95\% HDI for the corresponding models.}
    \label{fig:real data model fit assessment}
\end{figure}

Note however, that despite this the censored-GEV model still gives only a small probability ($2.2$\%) of a positive shape parameter $\Pr\{\xi>0\}=0.022$, that is, the model implies there is very likely to be an upper limit to the maximum possible pit depth. Our experience is that in most cases we estimate a negative shape parameter for maximum wall loss, in accordance with the findings in \cite{babu2019analysis}.
The QQ-plot in Figure~\ref{fig:real data model fit assessment} (a) indicates the fit quality of both the censored-GEV model (blue) and the GEV model (orange), with 95\% confidence intervals given as whiskers. Both models indicate a good fit over their domain, that is, the GEV model fits well over the whole domain, whilst the censored-GEV model fits well for those data points over the threshold. We also check the fit via the posterior predictive distribution for each model, Figure~\ref{fig:real data model fit assessment} (b). We see here that the GEV model has a good coverage of the observed data for those data points below the threshold. The fit of our model below the threshold is only for illustrative purposes, but the precise values are not of importance since we censor observations below $u=0.11$ (dashed black line). The upper tails of the two models appear similar and fit the data well, and indeed when comparing the upper 95\% highest density interval (HDI) (vertical lines), we see that the models give similar quantiles. 

\subsection{Maximum wall loss prediction}
\begin{figure}[h!]
    \centering
    \includegraphics[width=0.5\linewidth]{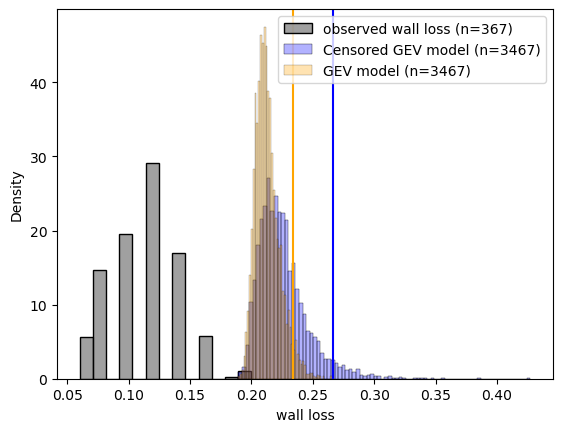}
    \caption{Posterior distribution of maximum wall loss over all unobserved tubes for both models, with corresponding vertical lines indicating the models' upper 95\% HDI.}
    \label{fig:real data max pred}
\end{figure}

Using the model fits above we now extrapolate to determine the distribution of the maximum wall loss over the $n^*$ unobserved tubes. For each MCMC sample (from 8000 samples), $n^*$ pit depths are drawn from a GEV distribution with location, scale and shape parameters corresponding to that sample. Of these $n^*$ tubes, the maximum value is stored, providing a distribution of maximum wall loss estimate for the two models. Figure~\ref{fig:real data max pred} shows the observed wall loss data (grey) with the censored-GEV model (blue) and GEV model (orange) predictions for maximum wall loss over the remaining unobserved $n^*=3489$ tubes. The vertical blue and orange lines indicate the corresponding 95th upper HDI of the models. 

Even though the models appear to fit similarly according to their posterior predictive distributions, Figure\ref{fig:real data model fit assessment}, the predictions for maximum wall loss over all tubes can look quite different, Figure\ref{fig:real data max pred}, with this difference likely attributable to the difference in shape parameters.

\section{Discussion}\label{sec:Discussion}
By developing a block maxima over threshold approach, we derive a censored-GEV framework to model wall loss that is still able to leverage the underlying limit theory of extremes even when the usual requirements for both block maxima or peaks over threshold cannot be assumed, resulting in a model which is more robust to model misspecification than the standard GEV approach.
Further, the Bayesian framework we propose brings multiple benefits over the typical frequentist approach used in applications of EVT to corrosion science:
\begin{itemize}
    \item The Bayesian framework facilitates direct inclusion of prior knowledge of corrosion levels either based on high availability of data or expert insight on typical corrosion behaviour in the given tube conditions, not possible using a frequentist approach. Note that non-informative priors are used in this paper because the focus of the simulation study here is on model-comparison between censored-GEV and GEV.
    \item Where an additional step of bootstrapping is needed to derive uncertainty estimates for a frequentist approach, with a Bayesian approach uncertainty estimates result from the analysis 'for free'. Bootstrapping is also often not suitable for extreme value theory as discussed earlier, so a Bayesian approach is needed to obtain uncertainty estimates.
\end{itemize}

We find in the simulated example in Section \ref{sec:sim study} that our censored-GEV model predicts return levels from a mixture distribution that are consistent with simulated data from a GEV, whereas the standard approach underestimates levels of corrosion, and does so with increased confidence. The real data example in Section \ref{sec:real data} agrees with the findings from the simulation study, with the standard approach predicting a smaller mean level of maximum corrosion and with increased certainty over our censored GEV-model. These results indicate the need for a censored-GEV approach, as using the standard GEV model may underestimate levels of corrosion if the model is fit to low wall loss samples that follow a different distribution from deep pits. Thus we have demonstrated the censored-GEV models increased confidence in the probable pitting corrosion across a heat exchanger for which only a subsample of tubes can be inspected, which supports timely maintenance of equipment critical to cooling in a processing line.

The primary concern in inspecting heat exchangers is indeed one of risk management - maintaining the integrity of tubes in a heat exchanger - which motivates consideration of the maximum pit depth (i.e. worst observed corrosion along the tube). But within the Bayesian paradigm we can make additional inference by simulating from the generative model. For example, if we expect pits of depths greater than $x$ to start impacting performance heavily, then we can answer questions such as, ``what is the total number of pits $N_x$ with depth greater than $x$ in the unobserved tubes?'' Here, with $x=0.18$ for example, then the mean and 95\% HDI are $\mathbb{E}[N_x] = 329 (241, 419)$. If this number is deemed too large then it might be deemed worthwhile to be more conservative and replace the tubes early even if the \textit{worst} pit depth does not pose an immediate risk. This extra level of inference therefore opens up the opportunity to use EVT to aid in joint decision making around questions of optimal performance as well as risk management.

\appendix
\renewcommand{\theequation}{A.\arabic{equation}}
\renewcommand{\thefigure}{A.\arabic{figure}}
\renewcommand{\thetable}{A.\arabic{table}}

\section{Additional figures from simulation study}
\label{sec: appendix additional sim figures}
The figures below contain additional results for the simulation study of bias and ratio of std. dev. for the location and scale parameters.
\begin{figure}[ht]
    \centering
    \begin{subfigure}{0.45\textwidth}
        \includegraphics[width=\textwidth]{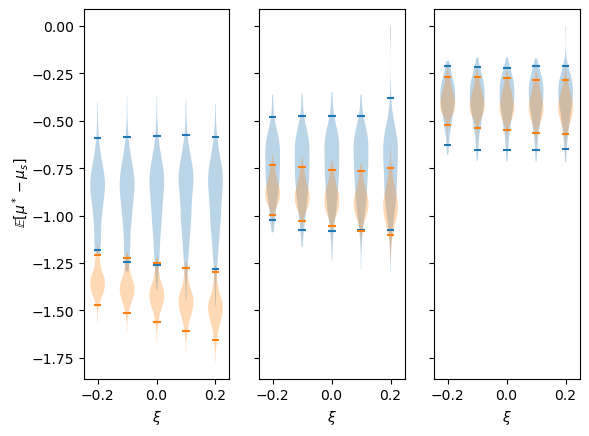}
        \caption{Bias}
    \end{subfigure}
    \hfill
    \begin{subfigure}{0.45\textwidth}
        \includegraphics[width=\textwidth]{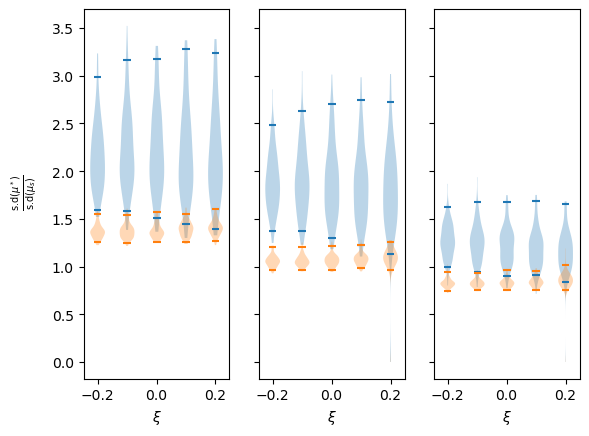}
        \caption{Std. dev. ratio}
    \end{subfigure}
    \caption{Posterior distribution of the \textbf{location} parameter bias (a) and ratio of std. dev. (b) compared to the simulated model, with $95\%$ HDI bars, for our censored-GEV model (blue) and GEV model (orange).}
    \label{fig:sim results location parameter}
\end{figure}

\begin{figure}[ht]
    \centering
    \begin{subfigure}{0.45\textwidth}
        \includegraphics[width=\textwidth]{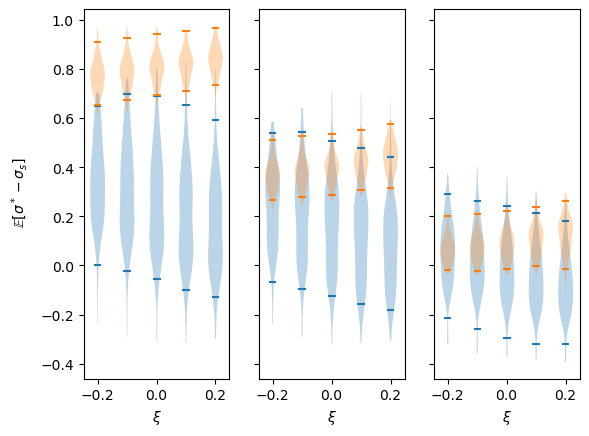}
        \caption{Bias}
    \end{subfigure}
    \hfill
    \begin{subfigure}{0.45\textwidth}
        \includegraphics[width=\textwidth]{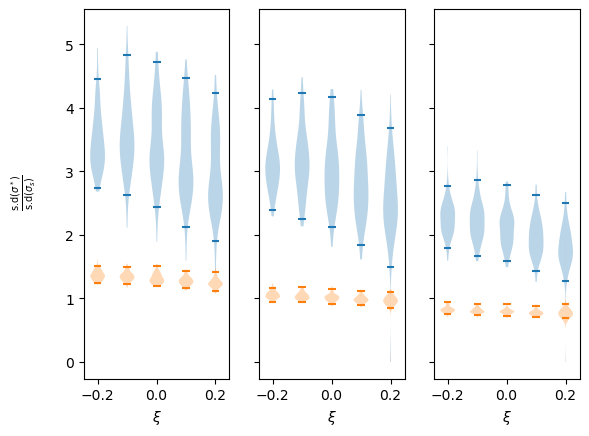}
        \caption{Std. dev. ratio}
    \end{subfigure}
    \caption{Posterior distribution of the \textbf{scale} parameter bias (a) and ratio of std. dev. (b) compared to the simulated model, with $95\%$ HDI bars, for our censored-GEV model (blue) and GEV model (orange).}
    \label{fig:sim results scale parameter}
\end{figure}


\clearpage
\bibliographystyle{apalike}
\bibliography{references}

\end{document}